\documentclass{article}

\usepackage[T1]{fontenc}
\usepackage{graphicx}
\usepackage{amsmath,amssymb}
\usepackage{xcolor}
\usepackage[margin=1.5in,bottom=1in,top=1in]{geometry}
\usepackage{float}
\usepackage[section]{placeins}
\usepackage[round,authoryear]{natbib}
\usepackage[colorlinks=true, linkcolor=blue, citecolor=blue, urlcolor=blue]{hyperref}
\usepackage[title,header,page]{appendix}

\usepackage[affil-it]{authblk}
\usepackage{lineno}
\usepackage{caption}
\usepackage{titlesec}
\titleformat{\section}[block]{\filcenter}{\thesection.}{1em}{\MakeUppercase}
\titleformat{\subsection}[hang]{\filcenter}{\thesubsection.}{1em}{\it}
\titleformat{\subsubsection}[hang]{\filcenter}{\thesubsubsection.}{1em}{\it}

\makeatletter
\renewcommand{\maketitle}{\begin{center}
  {\@title \par}
  \vskip 1em 
  {\large \@author \par}
  \vskip 0.0em 
  {\normalsize \@date \par}
  \end{center}
}
\makeatother

\usepackage{multicol}
\graphicspath{{figures/}}

\usepackage{listings}
\usepackage{needspace}
\newcommand{\kvec}{\mathbf{k}}
\newcommand{\xvec}{\mathbf{x}}
\newcommand{\avg}[1]{\left\langle #1 \right\rangle}
\newcommand{\gest}{\ensuremath{\mathbf{e}}}
\newcommand{\Rmatrix}{\ensuremath{\mathbf{R}}}
\newcommand{\code}[1]{\texttt{#1}}
\newcommand{\fixnoise}{\code{fixnoise}}

\newcommand{\mcal}{\textsc{metacalibration}}
\newcommand{\mdet}{\textsc{metadetection}}
\newcommand{\mcalshort}{\textsc{metacal}}
\newcommand{\Mcal}{\textsc{Metacalibration}}

\newcommand{\fusion}{\code{fusion}}
\newcommand{\noshear}{\code{noshear}}
\newcommand{\galsim}{\textsc{galsim}}
\newcommand{\noisevarfactor}{\ensuremath{F_V}}

\begin{document}

\title{\large \uppercase{A High Precision Metacalibration Response Correction}}
\date{\small \today}
\author{\small Erin S. Sheldon}
\affil{\footnotesize Brookhaven National Laboratory, Bldg 510, Upton, NY 11973, USA}
\maketitle

\renewcommand{\abstractname}{\normalfont ABSTRACT}

\begin{abstract}
    In \Mcal{} the shear response of a weak lensing measurement is estimated by
    applying an artificial shear to the image, but the shear operation imprints
    a spin-2 modulation on the image noise power, which biases the measured
    response. The standard fix adds a sheared noise field
    that symmetrizes the anisotropy, at the cost of doubling the noise
    variance ($\sqrt2$ in the noise). The \fusion{} correction presented in this
    note reaches the same unbiased response while adding only the minimum power
    needed. It measures the noise power transfer by passing an impulse through
    the \mcalshort\ process,
    forms the common-level azimuthal deficit that equalizes the spin-2
    ($m{=}2$) and spin-6 ($m{=}6$) modulations across the metacal types, and
    adds just that deficit by filtering the rotated noise field.
    The \fusion{} correction raises the per-pixel noise by only
    $\sim2\%$ for a round PSF (versus $\sqrt2$ for the standard correction),
    rising toward the $\sqrt2$ level only for highly elliptical PSFs.
\end{abstract}

\section{Overview}
\label{sec:overview}

\Mcal{} measures the response of an estimator to a small applied shear by
literally applying it \citep{huff2017metacal,sheldon2017metacal}. An observed
image is deconvolved by the point spread function (PSF), sheared by $\pm$ a
small step in each component, and reconvolved by a slightly larger round target
PSF. Differencing a shear estimator \gest\ from the sheared images gives the
response matrix \Rmatrix, and the calibrated shear is $\mathbf{g} =
\avg{\Rmatrix}^{-1} \avg{\gest}$.

The difficulty addressed in this note is noise. The deconvolution divides by
the PSF in Fourier space, which amplifies the high-$k$ noise; the shear
then stretches that amplified noise anisotropically. The \mcalshort'd image
therefore has noise with a power $\mathcal{P}(\kvec)$ that is no longer
isotropic but modulated as $\cos 2\theta$ (spin-2), and the modulation differs
between the \mcalshort{} types ($\pm$ and \noshear{}). Because the response is
a spin-2 measurement, a type-dependent spin-2 noise anisotropy leaks
into \Rmatrix\ and biases the recovered shear
\citep{GurvichMandelbaum2016,sheldon2017metacal}.

Two ways to remove it are
\begin{description}
\item[\fixnoise{} (exact, noisy).] Take an independent noise field,
    rotate it by $90^\circ$, \mcalshort{} it, rotate back and
    add it to the image \citep{sheldon2017metacal}. A $90^\circ$ rotation flips
    the sign of every spin-2 (and spin-6) anisotropy, so the sum is
    isotropic in those harmonics. But the
    added field doubles the noise power, $\sqrt2$ in the noise, $\times2$
    in variance, in every mode, whether or not that mode needed
    filling.
\item[\fusion{} (this note).] Add to each $k$-mode only the power needed
    to cancel the spin-2 effect. Most of the \mcalshort{} noise anisotropy is
        small; \fixnoise{} fully doubles the power at all $k$ to correct it. The
        \fusion{}\footnote{The name \fusion{} comes from the 
        fusion of two ideas, \fixnoise{} and an earlier attempt to correct without adding
        any extra noise.} correction measures how much power each
        \mcalshort{} type lacks relative to the others (the deficit), and
        supplies just that, by filtering the rotated field down to the deficit
        instead of adding it whole. The response is corrected to the same level
        as \fixnoise{}, but noise increase is $\sim\!1.02$ instead of $\sqrt2$
        for a round PSF\footnote{The minimum increase depends on the applied artificial
        shear; 1.02 is for an applied shear step of $\pm0.01$ and a dilation factor
        of $1 + 2\, \texttt{step}$}.
\end{description}

\subsection{The Flow}

The structure of this note follows the algorithm flow, which is:
\begin{enumerate}
\item \textbf{$k$-space metacal} (\S\ref{sec:metacal}): the operator
    $K_t$ for each \mcalshort{} type $t$ that deconvolves, shears, and reconvolves a field in
        \galsim\ $k$-space \citep{galsim2015}.
\item \textbf{The impulse transfer} (\S\ref{sec:transfer}): push an impulse
  through $K_t$ to produce the per-type noise power transfer
  $\mathcal{P}_t = |K_t|^2$, and see the type-dependent $m{=}2$/$m{=}6$ modulation.
\item \textbf{The harmonics that bias shear} (\S\ref{sec:harmonics}): the response
  isolates the $90^\circ$-odd harmonics $m{=}2,6$; $m{=}4$ is inert.
\item \textbf{The common-level deficit} (\S\ref{sec:deficit}): the power $D_t$
  that fills each type's mean and $m{=}2$/$m{=}6$ modulation up to the common
  (max-over-types) level, so every type gets one isotropic level.
\item \textbf{The rotated correction field} (\S\ref{sec:rotfield}): the source
  of the added power in the sky frame.
\item \textbf{The fusion filter} (\S\ref{sec:filter}): $H_t = \sqrt{D_t/
  \mathcal{P}_t^{\rm rot}}$, shaping the rotated field to the deficit.
\item \textbf{Assembly and output noise} (\S\S\ref{sec:assembly}--\ref{sec:noisevar}):
  one shared $k$-grid, and the predicted variance increase \noisevarfactor.
\end{enumerate}
The end-to-end validation is in \S \ref{sec:validation}.

\section{Metacalibration in $k$-space}
\label{sec:metacal}

In this section we describe the basic process implemented in the \texttt{metacal}
software package\footnote{\url{https://github.com/esheldon/metacal}}.  The
same underlying operations are used in \mdet\ \citep{mdet2020}, the nickname for
a \mcal\ that includes the object detection phase.

Write the observed image as $I = (G * P)$, the true surface brightness $G$
convolved by the PSF $P$. Define
the metacal type $t \in \{\noshear, 1p, 1m, 2p, 2m\}$ to correspond
with a shear operator $S_t$ that is identity for \noshear\ and a $\pm\,\text{step}$
shear in $g_1$ for 1p/1m and $g_2$ for 2p/2m.
Then the \mcalshort{}'d image is
\begin{equation}
  I_t = \Big[\, S_t\big(I * P^{-1}\big)\,\Big] * P_{\rm tgt},
  \label{eq:metacal-op}
\end{equation}
i.e.\ deconvolve the PSF, apply the shear in the world (sky) frame, and
reconvolve by a common target $P_{\rm tgt}$. The target is supplied by the
user\footnote{In this note we use the \code{AZGauss} variant from the \texttt{metacal}
package, a round Gaussian pinned
just below the input PSF's $k$-profile at the scale where that profile falls to
a small fraction of its peak, with the pinning done on the azimuthally averaged
profile so the target is insensitive to PSF-image noise at S/N $\gtrsim 50-100$}.
Whatever the target, \mcalshort\ dilates it by $1+2\,\texttt{step}$ to suppress
modes exposed by the shearing process.

All of \eqref{eq:metacal-op} is done in \galsim\ $k$-space. The image and PSF
are wrapped as \galsim\ \code{InterpolatedImage}s (Lanczos-15 $x$-interpolant,
the metacal kernel); the PSF is deconvolved, the result sheared, and the round
target reconvolved; the single complex profile is drawn once with
\code{drawKImage} onto a padded $N_p\times N_p$ grid for which the $k$-spacing,
\code{ifftshift} reordering, and centering phase are chosen to match
\code{numpy}'s FFT layout. This is followed by one inverse FFT and a crop to
the central $N\times N$ to yield $I_t$.  The pad size $N_p$ is
\galsim{}'s automatic \code{drawFFT} size; it is exposed only so that a set of
\mcalshort{}'d images can be forced onto one shared grid
(\S\ref{sec:assembly}), which the correction requires.


\section{The noise transfer from an impulse}
\label{sec:transfer}

The \mcal{} operation \eqref{eq:metacal-op} is linear in the input field, so in
$k$-space it acts mode-by-mode as multiplication by a complex transfer
$K_t(\kvec)$:
\begin{equation}
  \widehat{I_t}(\kvec) = K_t(\kvec)\,\widehat I(\kvec), \qquad
  K_t(\kvec) = \frac{\widehat{P_{\rm tgt}}(\kvec)}
                    {\widehat{P}(S_t^{-1}\kvec)}\;(\text{Lanczos, shear Jacobian}).
  \label{eq:ktransfer}
\end{equation}
If the input is a stationary noise field with white (flat) power spectrum, the
output noise power is
\begin{equation}
  \mathcal{P}_t(\kvec) \;=\; |K_t(\kvec)|^2 ,
  \label{eq:transfer-def}
\end{equation}
the per-type noise power transfer.

\begin{figure}[htbp]
\centering
\includegraphics[width=\textwidth]{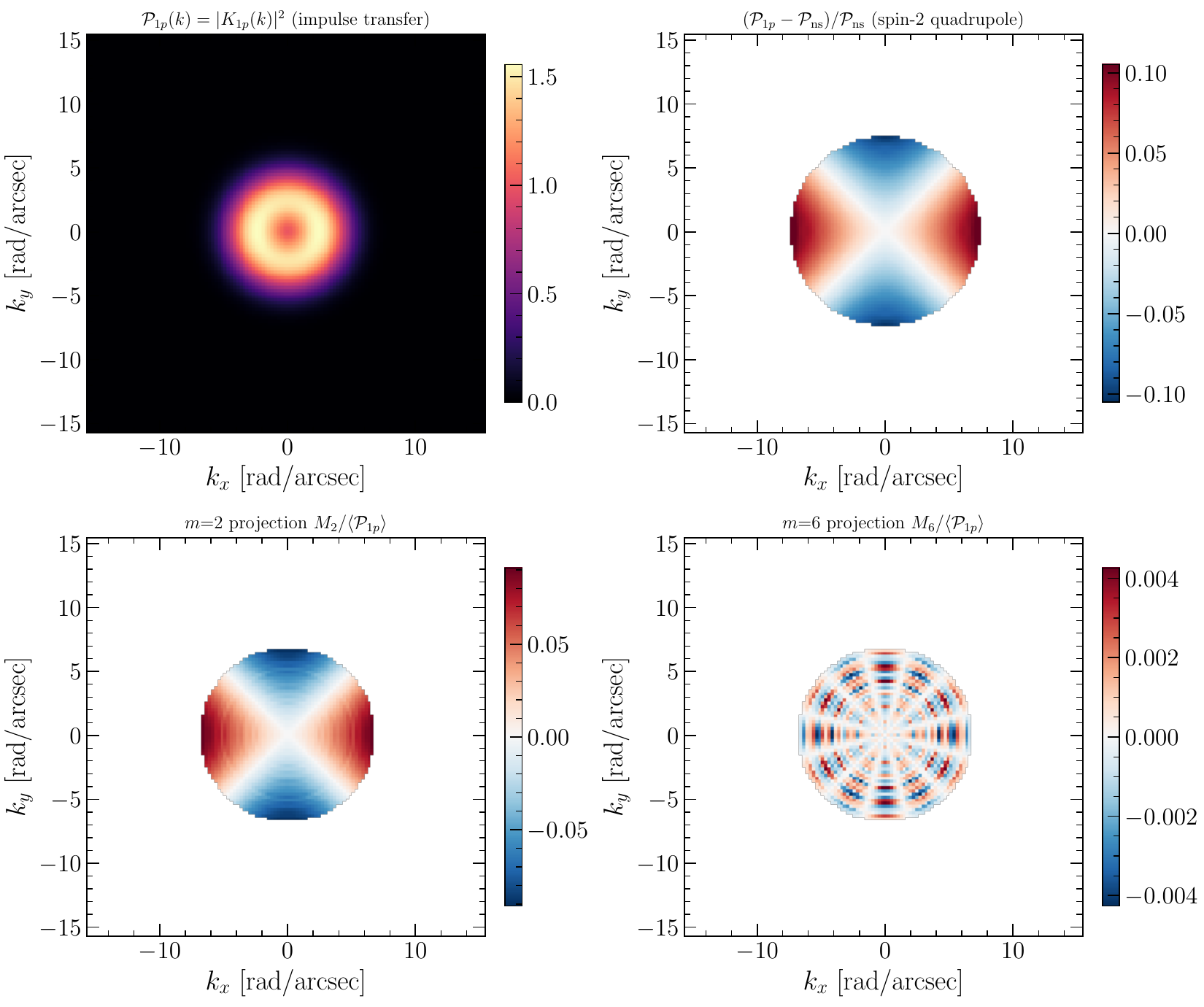}
\caption{The metacal noise transfer from a delta impulse
($\mathcal{P}_{1p}=|K_{1p}|^2$, round PSF, Moffat $\beta{=}2.5$ FWHM $0.9''$, diagonal
WCS, dim 48). \emph{Top-left:} the transfer, band-limited, peaking $\approx1.55$
in a ring. \emph{Top-right:} relative to \noshear, exposing the spin-2
quadrupole (enhanced along the $k_x$ stretch axis, suppressed along $k_y$).
\emph{Bottom:} the $m{=}2$ and $m{=}6$ azimuthal projections (note the
$\sim20\times$ smaller scale on $m{=}6$). These are the harmonics the
    \fusion{} filter fills.}
\label{fig:transfer}
\end{figure}

The form of $K_t$ in \eqref{eq:ktransfer} follows in two steps. First, a shear
is a linear coordinate remap of the image plane, and under Fourier transform it
stays a pointwise coordinate remap of
$\kvec$:
\begin{equation}
  \widehat{S_t f}(\kvec) = \widehat f(S_t^{-1}\kvec),
  \label{eq:shear-fourier}
\end{equation}
the inverse shear acting on $\kvec$,
with no amplitude factor because a pure
shear is area-preserving with a unit Jacobian. In practice the shear is applied in the sky
frame, so on the pixel $k$-grid the remap goes through the WCS
distortion matrix, but it remains a unit-determinant pointwise remap. Because
the remap acts pointwise it distributes over products and quotients, so
Fourier-transforming \eqref{eq:metacal-op} with
$\widehat{I * P^{-1}} = \widehat I/\widehat P$ gives
\begin{equation}
  \widehat{I_t}(\kvec)
  = \widehat{P_{\rm tgt}}(\kvec)\,
    \frac{\widehat{I}(S_t^{-1}\kvec)}{\widehat{P}(S_t^{-1}\kvec)} .
  \label{eq:sheared-quotient}
\end{equation}
The shear works on the numerator and the denominator. 
It must for a galaxy because shearing the galaxy is the point of \mcalshort. But
a stationary white noise field
has $\avg{|\widehat n(\kvec)|^2} = \sigma^2$, flat in $\kvec$, so the remapped
argument is unchanged. Sheared
white noise is statistically identical to white noise at second order. Taking
$\avg{|\cdot|^2}$ of \eqref{eq:sheared-quotient} therefore collapses all the
shear dependence of the noise power onto the deconvolution kernel, giving
\eqref{eq:transfer-def} with $K_t$ as in \eqref{eq:ktransfer}. Shearing the
deconvolved noise is statistically equivalent to deconvolving unsheared noise
by the sheared PSF, $\widehat{S_t P}(\kvec) = \widehat P(S_t^{-1}\kvec)$.
This is the sense in which the type-dependent anisotropy of $\mathcal{P}_t$ is
created by the \mcalshort\ shear (Figure~\ref{fig:transfer}). It enters
through the remapped argument $S_t^{-1}\kvec$, not through $\widehat P$, and
so flips sign between \code{1p} and \code{1m} even for a round PSF.

The operator also includes the response of the Lanczos interpolant with which
\mcalshort{} represents the image and the PSF (the parenthetical in
\eqref{eq:ktransfer}), and the padding, which is why we measure $K_t$ rather
than evaluate the quotient analytically.  This is distinct from any
interpolation used to build a coadd: that acted on the noise before the
\mcalshort\ operator, and is part of the input spectrum $N$, not $K_t$. See
\S \ref{sec:corrnoise} for the treatment of correlated noise.

We measure $\mathcal{P}_t$ directly by pushing a unit delta impulse through the
same \mcal{} process as the data: a delta has $|\widehat\delta(\kvec)| = 1$ in
every mode, so the output $k$-array is $K_t$ and its squared modulus is
$\mathcal{P}_t$.  This is exact for the operator, including the
Lanczos kernel and padding, not an analytic approximation.  It is measured per
\mcalshort\ type, capturing the small but crucial way the shear changes the
noise power.  The white-noise assumption is separate from the operator.  For a
correlated input with spectrum $N(\kvec)$, equation \ref{eq:sheared-quotient} gives an
output noise power $|K_t(\kvec)|^2\,N(S_t^{-1}\kvec)$, with the shear acting
on $N$ as well (\S\ref{sec:corrnoise}).


In figure~\ref{fig:transfer} we decompose the transfer for the \code{1p} type (a
$+g_1$ stretch) for a round PSF and a diagonal WCS. The transfer
(top-left) is band-limited and looks round by eye, because the metacal step is
only $0.01$. Dividing out the isotropic envelope (top-right, the relative
difference to \noshear) exposes a spin-2 quadrupole. Power is
enhanced along $k_x$ (the stretch axis) and suppressed along
$k_y$, reaching $\pm10\%$ at the band edge. The formal $m{=}2$ and $m{=}6$
azimuthal projections (bottom) isolate the two harmonics the correction targets;
$m{=}6$ is real but only $\sim\!5\%$ the size of $m{=}2$. This type-dependent
$m{=}2$ exists even for a round PSF. It is created by the metacal shear, which
is why it cannot be ignored.

\section{Which harmonics bias the shear}
\label{sec:harmonics}

Not every anisotropy of $\mathcal{P}_t$ matters. The metacal response in $g_1$ 
for estimator $e_1$ is
\begin{equation}
  R_{11} = \frac{e_1[I_{1p}] - e_1[I_{1m}]}{2\,\text{step}},
\end{equation}
which is a finite difference of two \mcalshort{} shears that are $90^\circ$ rotations of each other (a
$+g_1$ stretch along $k_x$ versus a $-g_1$ stretch along $k_y$). Hence \Rmatrix\ 
isolates the part of $\mathcal{P}_t$ that is odd under a $90^\circ$ rotation. A
harmonic $\cos m\theta$ satisfies
\begin{equation}
  \cos m(\theta + 90^\circ) = -\cos m\theta
  \quad\Longleftrightarrow\quad
  m \equiv 2 \pmod 4,
\end{equation}
i.e.\ $m = 2, 6, 10, \dots$ survive into \Rmatrix, while the even-under-$90^\circ$
harmonics $m = 0, 4, 8$ cancel. This is the reason the fusion fills
$m{=}2$ and $m{=}6$ and leaves $m{=}4$ alone: those are the
harmonics \fixnoise{} removes (adding the true $90^\circ$-rotated field zeroes
every $90^\circ$-odd harmonic), and $m{=}4$ is rotation-invariant and identical
across types, so it does not bias the response directly.

The $m{=}6$ part is spin-6, formally orthogonal to a
spin-2 estimator, yet it does bias the shear. The biasing quantity is the spin-2
moment of the weighted noise power,
$\int d^2k\, W(\kvec)\, e^{-2i\theta}\, \mathcal{P}(\kvec)$, with $W$ the
measurement kernel. Expanding both fields in harmonics, a noise harmonic $m$
produces a spin-2 only for a kernel harmonic $n = 2 - m$.
An isotropic kernel ($n{=}0$) admits only $m{=}2$; but the square pixel grid and
the Lanczos/fit weighting have an $m{=}4$ component ($n = \pm4$), which
demodulates $m{=}6$ down to $m{=}2$ through the identity
\begin{equation}
  \cos4\theta\,\cos6\theta = \tfrac12(\cos2\theta + \cos10\theta)
  \;\Longrightarrow\;
  \int_0^{2\pi}\!\cos2\theta\,\cos4\theta\,\cos6\theta\,d\theta = \tfrac{\pi}{2}\neq0 .
\end{equation}
So a spin-6 noise feature leaks into the shear by beating against a spin-4
anisotropy of the measurement. Filling $m{=}6$ removes that channel; an
end-to-end test (\S\ref{sec:validation}) measures its effect at $8$--$9\sigma$.

\section{The common-level harmonic deficit}
\label{sec:deficit}

We can only add noise power, so the correction will be a filtered noise field
with sufficient power to cancel the problematic $m{=}2/6$ modes. The fields
added to each \mcalshort{} type must have the same level in order for the
response to be correct.  Thus the goal is to raise every type up to a common
level, even \noshear{} which has no anisotropy for a round PSF. The common
level is the elementwise maximum over types, annulus by annulus, of the
azimuthal mean and of each filled $m$-mode amplitude. Filling each type up to
that envelope makes the corrected transfers identical across types and results
in isotropic total power in $m{=}2$/$m{=}6$.

For each
radial annulus we project every type's power onto its mean and its $(m{=}2,6)$
cosine/sine coefficients,
\begin{equation}
  \mu_t(k) = \avg{\mathcal{P}_t}, \qquad
  c^{(m)}_t(k) = \avg{\mathcal{P}_t\cos m\theta}, \qquad
  s^{(m)}_t(k) = \avg{\mathcal{P}_t\sin m\theta},
\end{equation}
form the common envelopes over types,
\begin{equation}
  \mu_\star(k) = \max_t \mu_t(k), \qquad
  A^{(m)}_\star(k) = \max_t \, 2\sqrt{c^{(m)2}_t + s^{(m)2}_t},
\end{equation}
and build the per-type deficit by (i) raising the mean to $\mu_\star$ and (ii)
for each $m$, cancelling that type's modulation and topping it up to the
common amplitude $A^{(m)}_\star$ (here aligned to the type's phase):
\begin{equation}
  D_t(\kvec) = \Big[\mu_\star - \mu_t\Big]
   + \sum_{m\in\{2,6\}} \Big[ A^{(m)}_\star
      - 2c^{(m)}_t\cos m\theta - 2s^{(m)}_t\sin m\theta \Big],
  \label{eq:deficit}
\end{equation}
clipped at zero (a power deficit cannot be negative).

\begin{figure}[htbp]
\centering
\includegraphics[width=\textwidth]{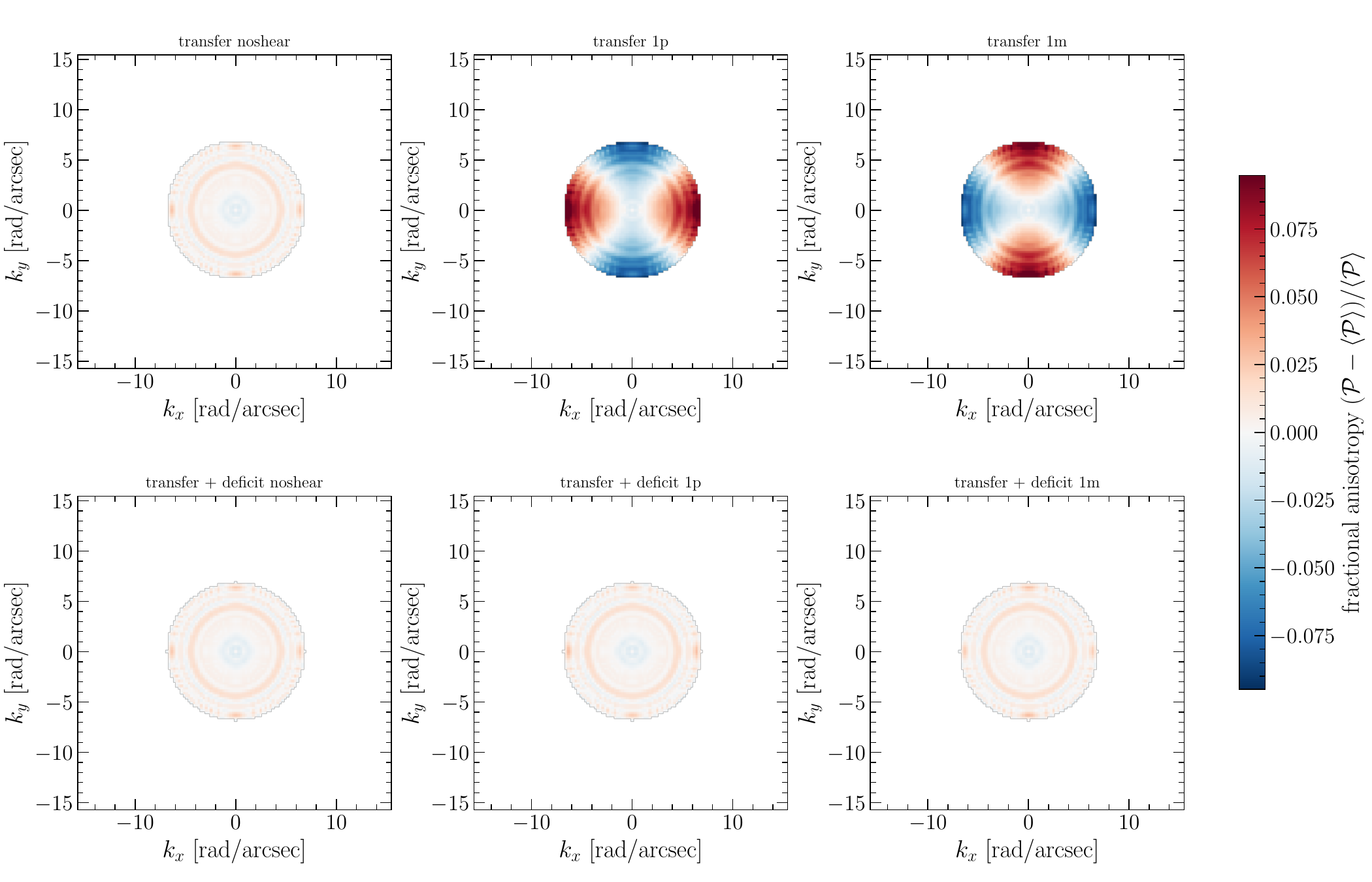}
\caption{Fractional
anisotropy $(\mathcal{P}-\avg{\mathcal{P}})/\avg{\mathcal{P}}$ (azimuthal mean divided out), shared colour
scale. \emph{Top:} the raw transfers. The metacal shear gives \code{1p} and
\code{1m} opposite quadrupoles even for this round PSF. \emph{Bottom:} after
adding the common-level $m{=}2/m{=}6$ deficit $D_t$, the three types are
identical and near-isotropic, so the response is unbiased. The faint residual is
the rotation-invariant $m{=}4$, deliberately uncorrected.}
\label{fig:isotropization}
\end{figure}

In figure~\ref{fig:isotropization} we show the effect. The raw per-type transfers
(top) have opposite quadrupoles. The \code{1p} is stretched along $k_x$
and the \code{1m} along $k_y$, with \noshear{} nearly round.  This is
the type-dependent spin-2 that biases
\Rmatrix. Adding $D_t$ (bottom) puts all three types on one near-isotropic level;
the faint residual ring is $m{=}4$ which does not produce a bias.

The azimuthal average $\avg{\cdot}$ neads to be calculated
carefully in order to avoid a bias, which we discuss in Appendices \ref{sec:trapz} and
\ref{sec:sky}.

\section{The rotated correction field}
\label{sec:rotfield}

The deficit $D_t$ gives how much power to add in each mode. The source of that
power is a rotated, \mcalshort{}'d noise field. Take an independent noise
realization with the same statistics as the data, rotate it by $90^\circ$ in
the sky frame, push it through the same \mcalshort{} operations, and rotate the
result back by $-90^\circ$. Its power transfer, measured by the same
delta-impulse trick is the rotated transfer $\mathcal{P}_t^{\rm rot}(\kvec) =
|K_t(\mathcal{R}_{90}\kvec)|^2$ where $\mathcal{R}$ is the rotation operator.

A $90^\circ$ rotation maps
$\cos2\theta \to -\cos2\theta$: it flips the sign of every spin-2 (and spin-6)
modulation. So where the image noise transfer $\mathcal{P}_t$ has $m{=}2$
troughs (along $k_y$ for a $+g_1$ stretch), the rotated transfer
$\mathcal{P}_t^{\rm rot}$ has $m{=}2$ peaks. These are the modes the deficit
$D_t$ is used to fill. The rotated field is thus naturally aligned with the deficit.
\fixnoise{} adds all of it (doubling the variance), while the fusion keeps only the
fraction that matches $D_t$. This alignment is visible in
Figure~\ref{fig:filter} below, where $\mathcal{P}_{\rm ns}^{\rm rot}$ and $D_{\rm ns}$
share the same lobe orientation.

For a conformal WCS a sky $90^\circ$ and a pixel rotation (e.g.
\code{numpy.rot90}) coincide. Under a non-conformal WCS they do not. A
pixel-grid rotation is not a sky $90^\circ$, so it fails to flip the sky-frame
spin-2 and leaves a spurious additive shear $c$. Rotating in the sky
frame, before metacal, then rotating back removes it. 

\section{The Fusion Filter}
\label{sec:filter}

We now have, on one shared grid, the un-rotated transfer $\mathcal{P}_t$ (the noise to
isotropize, with its deficit $D_t$) and the rotated transfer $\mathcal{P}_t^{\rm rot}$
(the correction field's power, in the same final frame after the rotation back).
The correction field is added through a filter $H_t(\kvec)$ applied to its
$k$-amplitude, so it contributes power $H_t^2\, \mathcal{P}_t^{\rm rot}$. Setting that
equal to the deficit, $H_t^2 \mathcal{P}_t^{\rm rot} = D_t$, gives
\begin{equation}
  H_t(\kvec) = \sqrt{\frac{D_t(\kvec)}{\mathcal{P}_t^{\rm rot}(\kvec)}}
  \label{eq:filtershort}
\end{equation}
Adding $H_t\cdot(\text{rotated
field})$ then gives a total noise power $\mathcal{P}_t + H_t^2 \mathcal{P}_t^{\rm rot} = \mathcal{P}_t + D_t$
on the common isotropic level of \S\ref{sec:deficit}.

In practice we need to be careful not to add more than the full
rotated field in any mode, and roll $H$ smoothly to zero:
\begin{equation}
  H_t(\kvec) = \min\!\left(\sqrt{\frac{D_t(\kvec)}{\mathcal{P}_t^{\rm rot}(\kvec)}},\ 1\right)
  \cdot \underbrace{\frac{\mathcal{P}_t^{\rm rot}}{\mathcal{P}_t^{\rm rot} + \epsilon}}_{\text{taper}}
  \label{eq:filter}
\end{equation}
with $\epsilon = \texttt{ktol}\cdot
\max \mathcal{P}_t^{\rm rot}$; we find $\texttt{ktol}=10^{-4}$ works well. 

The cap $\min(\cdot,1)$;  $H_t \le 1$, guarantees the \fusion\ correction never adds more
than the full rotated field in any mode. It can never add more noise
than \fixnoise{}. In practice the raw $\sqrt{D_t/\mathcal{P}_t^{\rm rot}}$ exceeds
$1$ only near the band edge (where $\mathcal{P}_t^{\rm rot}$ is collapsing), so
for a round or mildly elliptical PSF the cap is essentially inactive in-band;
it is a safety bound that engages only as the PSF becomes highly elliptical.

Out of band both $D_t$ and $\mathcal{P}_t^{\rm rot}$ go to zero, so
$\sqrt{D_t/\mathcal{P}_t^{\rm rot}}$ is a noisy $0/0$; the taper
$\mathcal{P}^{\rm rot}/(\mathcal{P}^{\rm rot}+\epsilon)$ rolls $H_t$ smoothly
to zero wherever $\mathcal{P}_t^{\rm rot}$ falls below $\epsilon$, preventing
the filter from amplifying numerical noise where there is no signal.

In figure~\ref{fig:filter} we show the construction for an elliptical PSF
with $g_1^{\rm psf}=0.01$ (reduced shear convention), comparable to the metacal step. The rotated
transfer
$\mathcal{P}_{\rm ns}^{\rm rot}$ (top-left) and the deficit $D_{\rm ns}$ (top-right) share
the same lobe orientation, confirming the alignment argument of
\S\ref{sec:rotfield}; the filter $H_{\rm ns}$ (bottom-left) carves the rotated
power down to the deficit. The $k$ cross-section (bottom-right) demonstrates the cap and taper.
The raw $\sqrt{D/\mathcal{P}^{\rm rot}}$ rises toward the band edge but
stays below the cap at $1$. The cap engages only for a highly elliptical
PSF (see Figure~\ref{fig:filter-elliptical}). The taper pulls the final $H$ to
zero beyond the band.

\newcommand{\fusionsize}{1.0}
\begin{figure}[h]
\centering
\includegraphics[width=\fusionsize\textwidth]{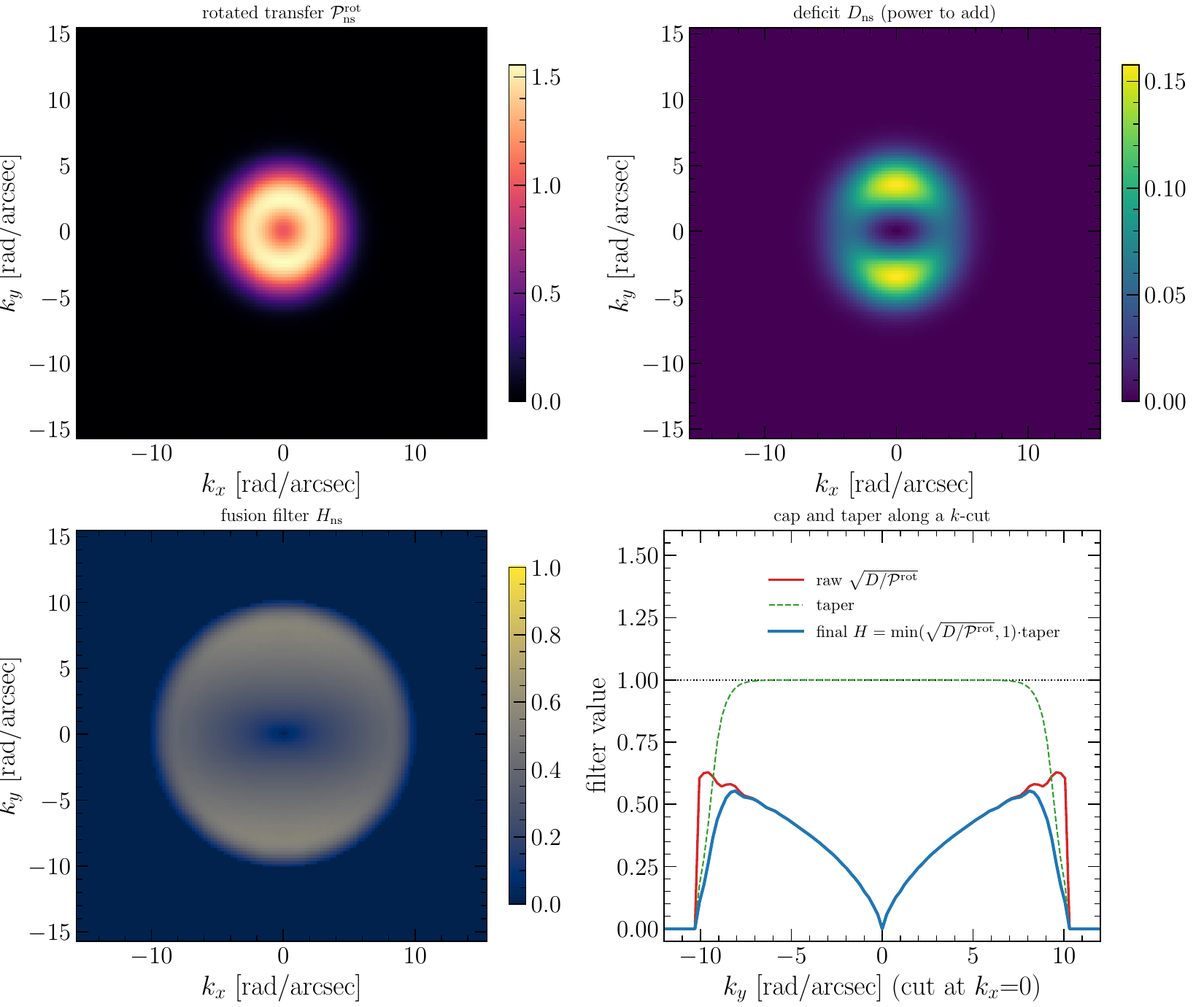}
    \caption{A \fusion{} filter for $\noshear$ type, $g_1^{\rm psf}=0.01$.
    \emph{Top-left:} the rotated transfer $\mathcal{P}_{\rm ns}^{\rm
rot}$, the correction field's power, lobed along $k_y$. \emph{Top-right:} the
deficit $D_{\rm ns}$ to add, with the same lobe orientation, the alignment that lets a
rotated field fill the deficit. \emph{Bottom-left:} the filter $H_{\rm ns}=
\min(\sqrt{D/\mathcal{P}^{\rm rot}},1)\cdot\text{taper}$. \emph{Bottom-right:} a $k$
    cross-section at
$k_x{=}0$: the raw $\sqrt{D/\mathcal{P}^{\rm rot}}$ (red) rises near the band edge
but stays below the cap at $1$, the taper (green) rolls off out of band, and the
final $H$ (blue) is their product.}
\label{fig:filter}
\end{figure}

In figure~\ref{fig:filter-elliptical} we show the same construction for a highly
elliptical PSF $g_1^{\rm psf}=0.15$. The
deficit is now of order the rotated transfer, the raw
$\sqrt{D/\mathcal{P}^{\rm rot}}$ reaches $1$ over half the band, and the
$\min(\cdot,1)$ saturates the filter at $H{=}1$. In those modes the \fusion\
correction adds the full rotated field, approaching \fixnoise{}. This is the
regime behind the rising added noise shown in \S\ref{sec:noisevar}.

\begin{figure}[h]
\centering
\includegraphics[width=\fusionsize\textwidth]{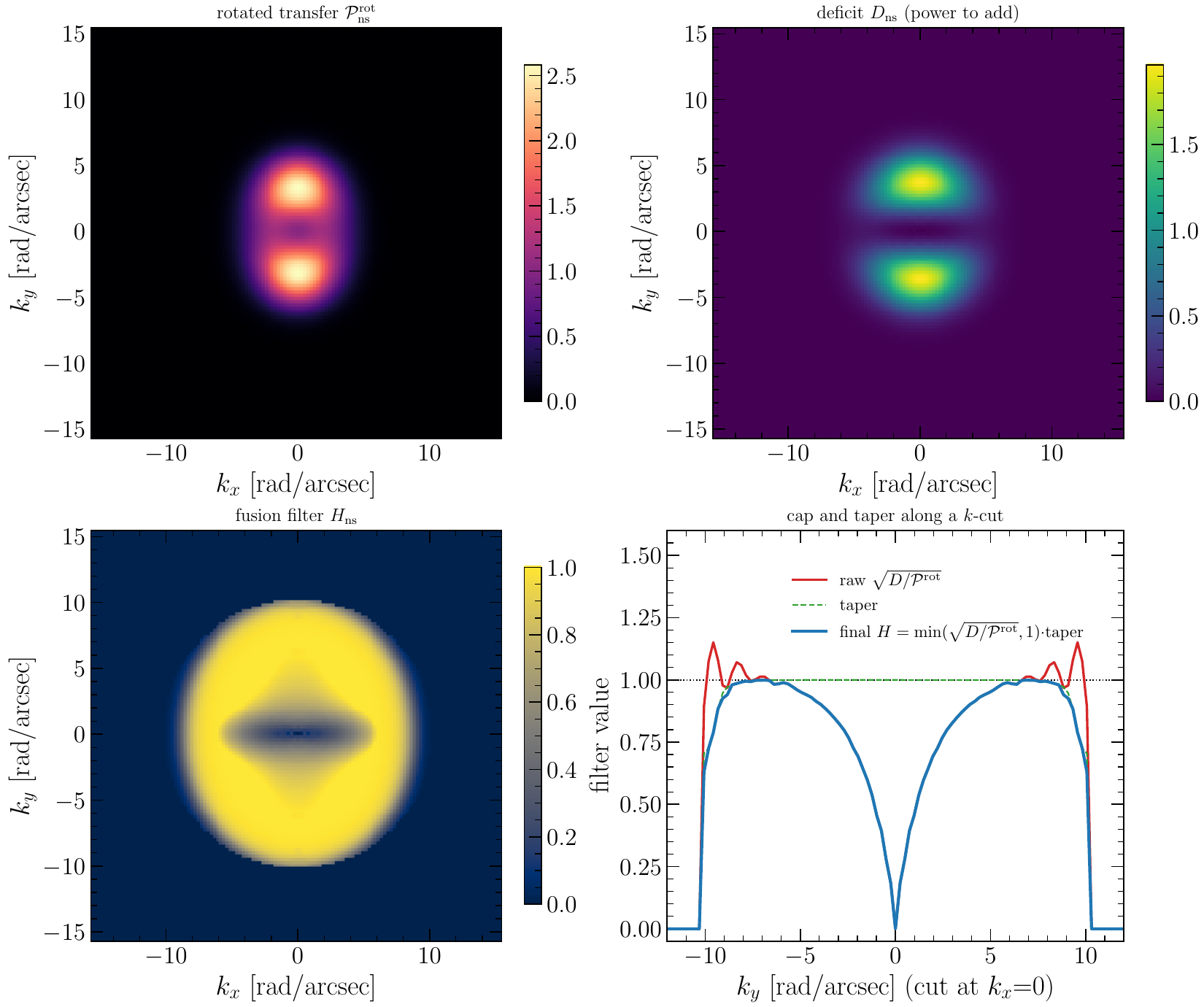}
\caption{Cap engagement for a highly elliptical PSF, with $\noshear$
type, $g_1^{\rm psf}=0.15$.  The layout is the same as as Figure~\ref{fig:filter}.
The deficit (top-right) is now comparable to the rotated transfer (top-left);
the filter (bottom-left) saturates at $H{=}1$ over half the band. In those
modes the fusion adds the full rotated field, as \fixnoise{} does everywhere.
On the $k_x{=}0$ cross-section the raw $\sqrt{D/\mathcal{P}^{\rm rot}}$ (red) crosses $1$
    and the final $H$ (blue) is at the cap.}
\label{fig:filter-elliptical}
\end{figure}

\begin{figure}[h]
\centering
\includegraphics[width=\fusionsize\textwidth]{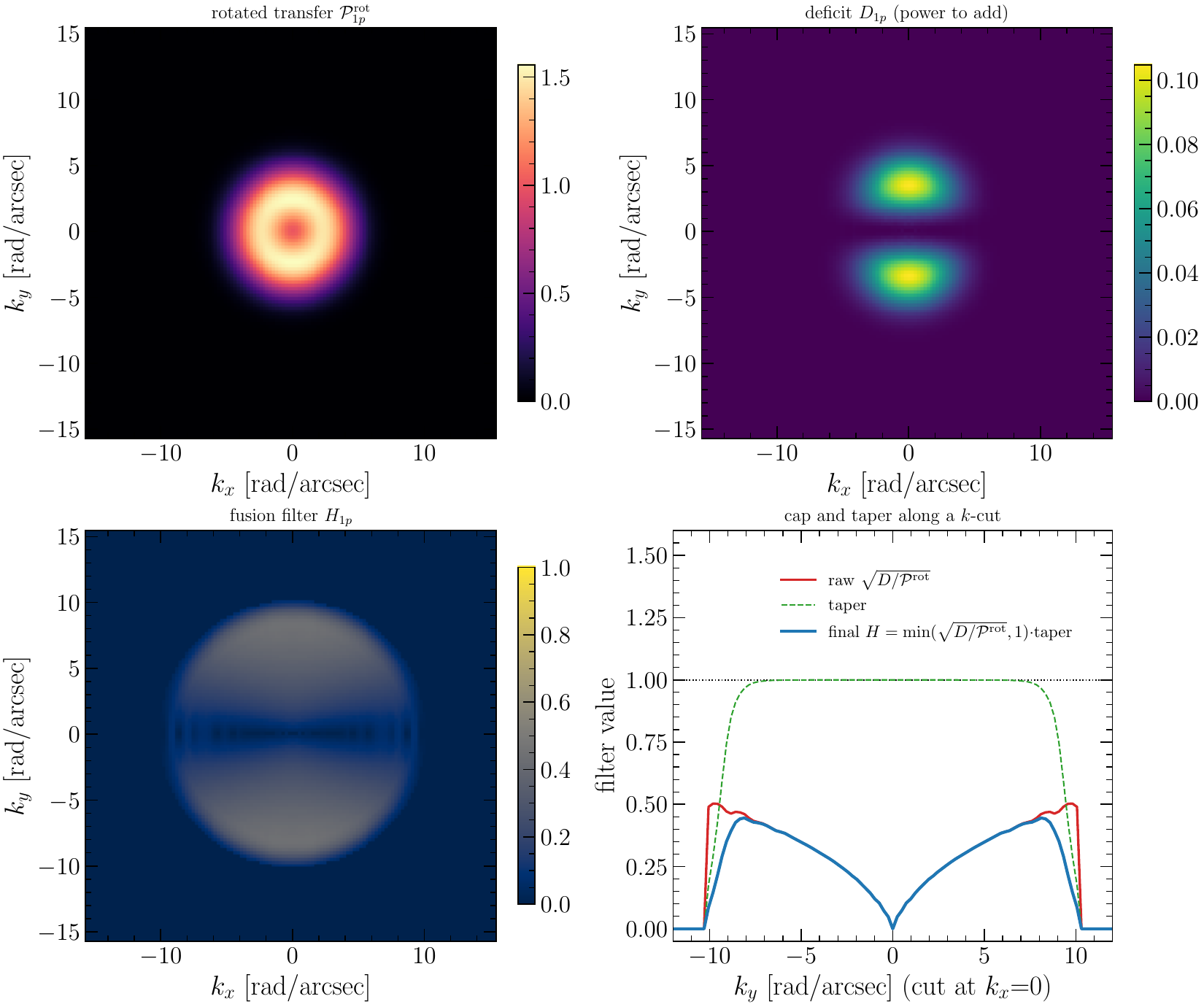}
\caption{The \fusion\ filter in the nominal case, with \code{1p} type, round
PSF, standard step $0.01$. The layout is the same as Figure~\ref{fig:filter}.
\emph{Top-left:} the rotated transfer $\mathcal{P}^{\rm rot}_{1p}$.
    \emph{Top-right:} the deficit $D_{1p}$, the two-lobed $m{=}2$
pattern along $k_y$ that fills the trough carved by the $+g_1$ stretch; note the
colour scale, $\sim\!7\%$ of the transfer peak. \emph{Bottom-left:} the filter
$H_{1p}$, near zero along $k_x$ where no power is missing. \emph{Bottom-right:}
the $k_x{=}0$ cross-section through the lobes. The raw $\sqrt{D/\mathcal{P}^{\rm rot}}$
(red) stays below the cap at $1$ everywhere in band, and the final $H$ (blue)
never exceeds $\sim\!0.43$.}
\label{fig:filter-round}
\end{figure}

In figure~\ref{fig:filter-round} we repeat the construction for the nominal case,
a round PSF, shown for the \code{1p} type, where the only anisotropy is the
standard metacal step ($0.01$). The deficit is the two-lobed
$m{=}2$ pattern along $k_y$, which is the trough the $+g_1$ stretch produces in
$\mathcal{P}_{1p}$ (\S\ref{sec:transfer}), topping out at $\sim\!7\%$ of the
transfer peak. The filter is correspondingly near zero along the $k_x$ axis.
No noise is added where the metacal shear already enhanced the power. On the
$k_x{=}0$ cross-section through the lobes the raw $\sqrt{D/\mathcal{P}^{\rm rot}}$ stays
below $\sim\!0.5$ everywhere, so the cap never engages; the added power
$H^2\mathcal{P}^{\rm rot}$ is a few percent of the plain metacal noise. This
is a geometrical explanatioin of the $\sim\!2\%$ round-PSF noise cost of
\S\ref{sec:noisevar}.

\subsection{Correlated Input Noise} \label{sec:corrnoise}

$\mathcal{P}_t$ and $\mathcal{P}_t^{\rm rot}$ are defined relative to white
input noise, and $H_t$ is exact only for white noise.  For a correlated input
with spectrum $N(\kvec)$ the image noise power is
$\mathcal{P}_t(\kvec)\,N(S_t^{-1}\kvec)$ (\S\ref{sec:transfer}).  The
$90^\circ$ rotation leaves an isotropic $N$ unchanged, but the \mcalshort{}
shear does not.  To first order in the shear $g_1$ of type $t$, the remap is
$S_t^{-1}\kvec \approx \big(k_x(1-g_1),\,k_y(1+g_1)\big)$, so for isotropic
$N(k)$,
\begin{equation}
  |S_t^{-1}\kvec| \approx k\,(1 - g_1\cos 2\phi),
  \qquad
  N(S_t^{-1}\kvec) \approx N(k)\left[1 - g_1\,\frac{d\ln N}{d\ln k}\cos 2\phi\right],
  \label{eq:corrnoise-remap}
\end{equation}
with $\phi$ the azimuth of $\kvec$.  The remapped $N$ has a spin-2
modulation of fractional amplitude $g_1\,d\ln N/d\ln k$.  Since
$g_1 = \pm\,\text{step}$ for 1p/1m (and the same with $\sin 2\phi$ for 2p/2m),
the modulation flips sign between the paired types.  It is type-odd so
enters the shear response (\S\ref{sec:harmonics}), and the white
deficit $D_t$ does not include it.  The second-order terms are type-common and
cancel in $R$.

How much this matters depends on how much $N$ varies across the PSF band.  For
typical coadds, Lanczos-3 resampling at unit pixel ratio leaves $N$ flat to
$\sim3\%$ across the band.  A $k$-space calculation with the exact resampling
kernel of a 10-epoch dithered coadd, with and without rotational dithers, gives
a type-odd residual after the correction of $6\times10^{-4}$ of the noise
power, compared to $4\times10^{-4}$ for white noise ($1.2\%$ vs.\ $0.8\%$ of the
uncorrected
anisotropy), and based on focused simulations we expect a multiplicative
bias of $m\sim10^{-3}$ for
typical Lanczos-3 coadds (\S\ref{sec:validation}) below requirements
    for upcoming surveys \citep[e.g.][]{SRD}.

For strongly correlated noise the white filter is not adequate. For white noise
smoothed by a Gaussian with $\sigma$ = 1 pixel it leaves half the type-odd
anisotropy in place ($1.6\times10^{-2}$ of $3.2\times10^{-2}$).  The remedy
is to measure the transfers by pushing the noise correlation kernel $\sqrt
N$ through the operator in place of the delta, after which the deficit
machinery is exact again (residual $3\times10^{-4}$ in the same
calculation); the kernel must come from the known resampling, as the
autocorrelation of a single noise realization is far too noisy.  The real
noise realization is still needed in any case as it accurately represents
whatever non-stationary structure the data noise has
(Appendix~\ref{sec:synth}).

The predicted \noisevarfactor\ of \S\ref{sec:noisevar} weights the band by $N$;
for correlated noise it is approximate.

\section{Assembling the correction on one shared grid}
\label{sec:assembly}

The filter, the deficit, and the noise to be cancelled must all be in the same
$k$-frame. The filter is applied in $k$ on the padded grid with no second FFT, so
the de-aliased correction field is in the same frame as the constructed transfer.

So the final image for type $t$ is the plain \mcalshort{}'d data plus the filtered,
rotated, metacal'd noise:
\begin{equation}
  I_t^{\rm corr}(\xvec)
  = \underbrace{\mathcal F^{-1}\!\big[K_t\,\widehat I\big]}_{\text{plain metacal}}
  + \underbrace{\mathcal F^{-1}\!\big[H_t\,K_t^{\rm rot}\,\widehat{n}\big]}_{\text{the deficit fill}},
\end{equation}
each cropped to the central $N\times N$.  $\mathcal F^{-1}$ represents
the inverse Fourier transform\footnote{In the \texttt{metacal} package,
the result is returned as a
\code{MetacalResult}, which is keyed by type to the corrected images and also
holds the target PSF image and the output-noise factor of the next
section.}. 

\section{The output noise level}
\label{sec:noisevar}

The correction adds anisotropy-filling power, so the corrected per-pixel noise
is higher than plain metacal. By Parseval on the padded grid, the
per-pixel variance of a filtered white field is the mean of its squared filter.
The plain \mcalshort\ contributes $\avg{\mathcal{P}_t}$ and the correction adds
$\avg{H_t^2 \mathcal{P}_t^{\rm rot}}$.  The deficit gives
every type a common noise level, and only the $m{=}0$ (azimuthal-mean) part
sets the total variance (the $m{=}2,6$ fills integrate to zero over each
annulus). Calculating from \noshear{},
\begin{equation}
    \noisevarfactor
  = 1 + \frac{\avg{H_{\rm ns}^2\, \mathcal{P}_{\rm ns}^{\rm rot}}}{\avg{\mathcal{P}_{\rm ns}}} .
  \label{eq:nvf}
\end{equation}
Downstream this multiplies the noise variance used for
uncertainty calculations\footnote{Equivalently, divides the weight
map. In the \texttt{metacal} package \code{metacal\_obs}, the
ngmix \url{https://github.com/esheldon/ngmix} \code{Observation} wrapper, applies it
automatically}. It is $1$ when no correction is applied.

\begin{figure}[h]
\centering
\includegraphics[width=0.75\textwidth]{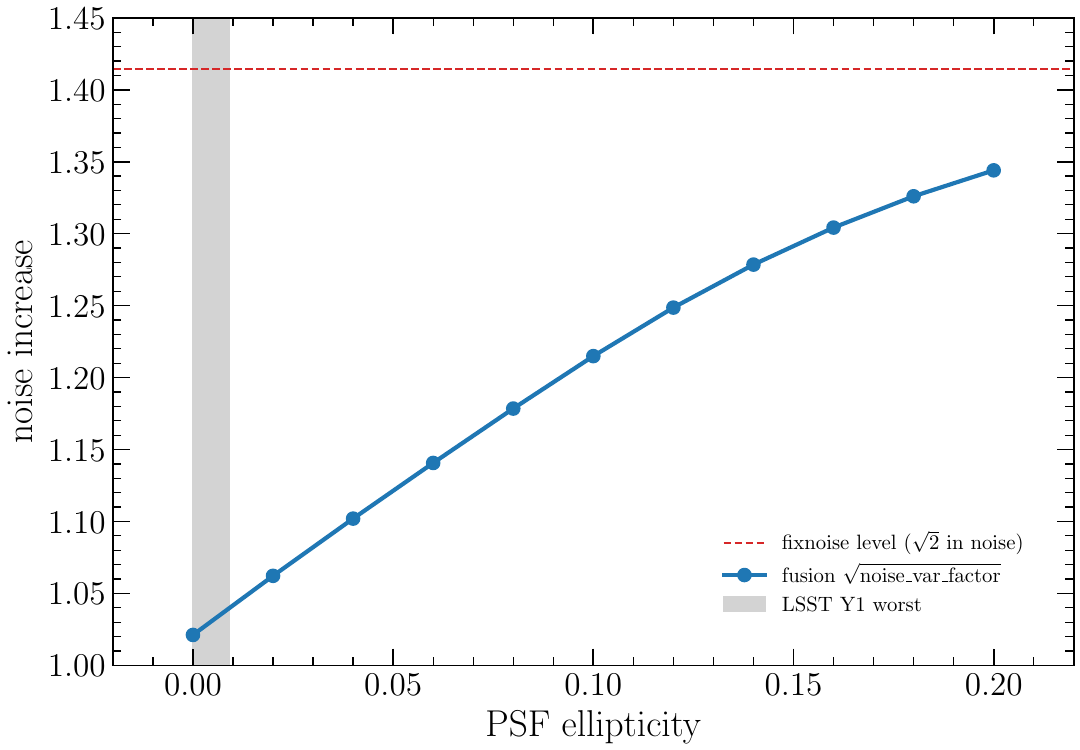}
    \caption{The increased noise standard deviation, relative
    to plain \mcal, of the \fusion{} correction vs.\ PSF ellipticity
    in the reduced shear convention.
    The \fusion{} increase is $\approx1.02$ for a round PSF and rises toward the
\fixnoise{} level of $\sqrt2$ (dashed) only as the PSF becomes highly
elliptical. \fixnoise{} increases by the noise by $\sqrt2$ ($\times2$ in variance)
in every case.}
\label{fig:cost}
\end{figure}

In figure \ref{fig:cost} we plot the resulting noise (std) increase
$\sqrt{\noisevarfactor}$ versus PSF ellipticity in the
reduced shear convention \citep{bj02}. For a round PSF the
variance factor is $\approx1.04$, which is only $\sim2\%$ in the noise, so the
correction is nearly free.  It rises toward \fixnoise's $\sqrt2$ in noise
($\times2$ in variance) only as the PSF becomes highly elliptical
($\sqrt{\noisevarfactor}\approx1.34$ at $g_1^{\rm psf}=0.2$).
Deconvolving by an elliptical PSF makes the metacal noise more anisotropic, so
more of the rotated field is needed to fill the deeper deficit, and the cost
approaches the full doubling of \fixnoise{}. The
prediction \eqref{eq:nvf} matches the measured ratio of corrected to plain noise
variance to better than $5\%$ in direct white-noise tests.

%

\section{End-to-end validation}
\label{sec:validation}

The pieces above were each validated in dedicated shear-bias simulations, based
on those from \citep{sheldon2023rubin} but with additional features to
support simulated coadds with edges
\citep{sheldon2026edges}\footnote{\url{https://github.com/esheldon/simcoadd}}.
For the primary simulations, we used , ``edgy''-coadd noise with 10 input
epochs
\citep{sheldon2026edges}, semi-realistic galaxy population
\citep{sheldon2023rubin,wldb2021}, a $1.1\%$ non-conformal
WCS plus a $10^\circ$ rotation, and a spatially variable, elliptical Moffat PSF
\citep{Moffat1969}.

For the analysis we used and full \mdet{} with $k$-space metacal,
\texttt{sep} detection \citep{sep} and a maximum-likelihood Gaussian fit.
We did not push a correlated noise kernel through the deficit machinery so, as
noted in \S \ref{sec:corrnoise}, for coadd correlated noise from
Lanzcos-3 resampling we expect $m \sim 0.001$.  The results were
\begin{itemize}
\item Multiplicative bias $0.0004 < m < 0.0014$
    at 99.7\% confidence, within LSST
    tolerances \citep{SRD} and consistent with the expectation
    for uncorrected correlated noise.
\item In a paired, non-noise-cancelling run,  we measured
  $\mathrm{var}^{\fixnoise}/\mathrm{var}^{\rm fusion} = 2.043\pm0.005$: \fixnoise{}
        costs essentially double the shear-noise variance of the \fusion{} 
        correction (the residual
        above $2$ is the \fusion{}'s $\sim2.6\%$ larger restored response, not extra image
        noise).
\item A matched-seed A/B
  ($m{=}\{2,6\}$ vs.\ $m{=}2$-only, $1.25\times10^6$ trials, $11\%$ non-conformal
  WCS, sky-rotated correction) resolves $m{=}6$ at $8$--$9\sigma$: dropping it
  leaves a $+3.5\times10^{-4}$ multiplicative bias that the $m{=}6$ fill removes,
  consistent with the \fixnoise{} reference.
\end{itemize}

\section{Summary}
\label{sec:summary}

The fusion noise correction makes the metacal noise spin-2 isotropic
while adding the minimum amount of noise power. The flow is:
\begin{enumerate}
    \item Push a unit delta impulse through the \mcal{} to get the per-type noise
  power transfer $\mathcal{P}_t = |K_t|^2$ (\S\ref{sec:transfer}).
\item Form the common-level deficit $D_t$ that fills each \mcalshort\ type's mean and
  $m{=}2$/$m{=}6$ modulation up to the max-over-types envelope.
\item Measure the rotated transfer $\mathcal{P}_t^{\rm rot}$ of a sky-$90^\circ$-rotated,
    \mcalshort{}'d noise field (\S\ref{sec:rotfield}).
\item Filter that field by
    $H_t = \sqrt{D_t/\mathcal{P}_t^{\rm rot}}$ with a cap and a taper
    and add it to the \mcalshort{}'d data on one shared $k$-grid
  (\S\S\ref{sec:filter}--\ref{sec:assembly}).
\end{enumerate}
The noise increase is \noisevarfactor\ $= 1 + \avg{H^2 \mathcal{P}^{\rm rot}}/\avg
P$, $\approx1.04$ for a round PSF and rising toward $2$ only for highly
elliptical PSFs (\S\ref{sec:noisevar}), compared to \fixnoise's
$\times2$ ($\sqrt2$ in the noise). The corrected estimator is
accurate to LSST requirements even under combined
correlated non-stationary noise, non-conformal WCS, and PSF ellipticity.

\section*{Acknowledgments}
We thank the computing staff of the Scientific Computing \& Data Facility at
Brookhaven National Laboratory for their support. 

\begin{appendices}

\section{The rotation-covariant azimuthal quadrature}
\label{sec:trapz}

On a square FFT grid an annulus is not a continuous circle but the finite set of
lattice modes $|\kvec|$ that round into each radial bin. Those modes
have the $D_4$ symmetry of the lattice.
The lowest surviving anisotropic moment is therefore
\begin{equation}
  Q \equiv \avg{\cos 4\theta} \neq 0,
\end{equation}
with $Q>0$ when modes cluster on the pixel axes, $Q<0$ on the diagonals.
This $Q$ bleeds into the discrete $m{=}2$ projection.
Projecting a pure $m{=}2$ pattern of amplitude $A$ at orientation $\varphi$,
$\mathcal{P} = A\cos2(\theta-\varphi)$, the integrand self-beats to a constant plus an
$m{=}4$ term, and the $m{=}4$ term does not integrate to zero on the
discrete ring because $\avg{\cos4\theta}=Q$:
\begin{equation}
  c_2 = \tfrac{A}{2}(1+Q)\cos2\varphi, \qquad
  s_2 = \tfrac{A}{2}(1-Q)\sin2\varphi.
\end{equation}

The two \mcalshort{} shears make $m{=}2$ deficits with eigenaxes that are
$45^\circ$ apart: $g_1$ (\code{1p}/\code{1m}) along the world axis, $g_2$
(\code{2p}/\code{2m}) along the diagonal. The biased projection therefore
over-fills one and under-fills the other, leaving a response anisotropy $\delta
= \tfrac12(R_{11}-R_{22}) \propto Q$. Rotating the WCS turns the world axes
against the fixed grid, so $\delta \propto \cos4\theta_w$, causing a rotation
leak.

We can replace the uniform discrete mean $\frac1N\sum_i$ by a weighted average
so that the spurious moments are reduced. We chose an arc-length (trapezoidal /
Voronoi) quadrature. Within each annulus, we sort the mode angles and weight each
mode by the angular gap it occupies, $w_i \propto
\tfrac12(\theta_{i+1}-\theta_{i-1})$ cyclically, normalized to $\sum_i w_i =
1$. This is the trapezoidal rule on the non-uniform angular grid. It
approximates the continuum integral $\frac1{2\pi}\int\cos p\theta\,d\theta = 0$
and results in $M_p = \avg{\cos p\theta}$ being reduced to the trapezoidal error
for $m{=}4,8,\dots$.  It is
pure grid geometry, so it is computed once per grid, with no noise.

\begin{figure}[htbp]
\centering
\includegraphics[width=0.75\textwidth]{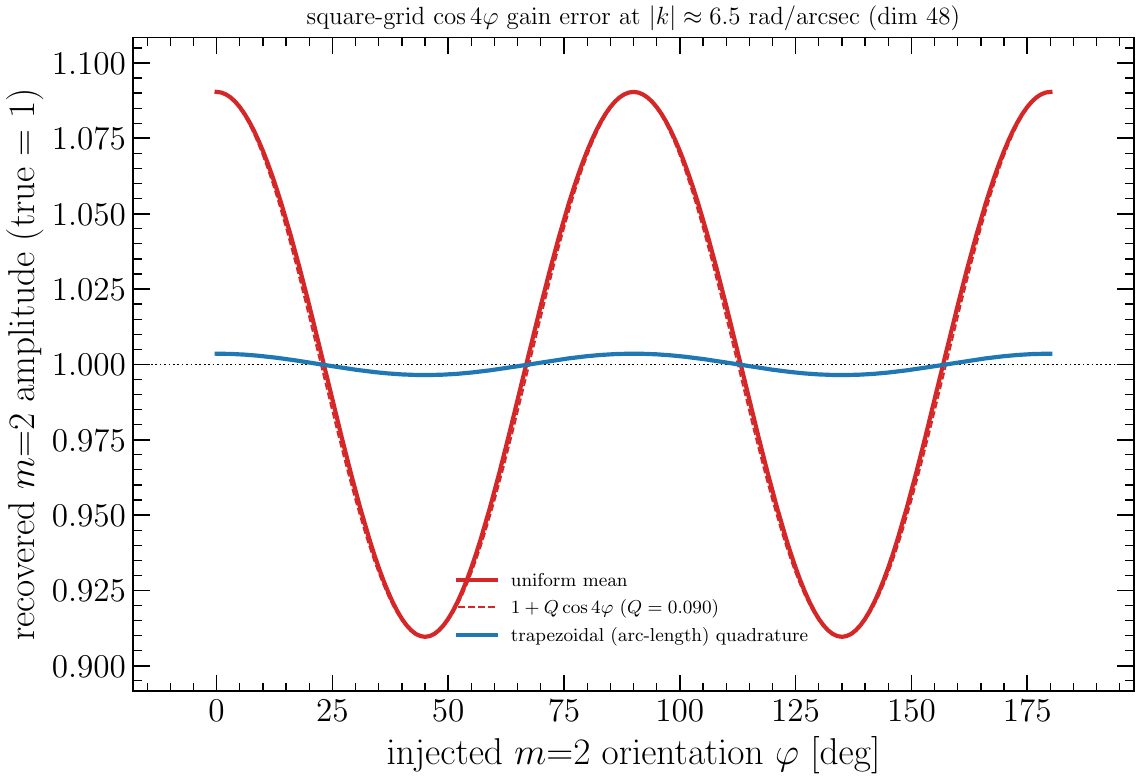}
\caption{
Recovered amplitude of a unit $m{=}2$ pattern vs.\ its orientation $\varphi$, on
an in-band annulus (dim 48). The uniform mean (red) ripples as
$1+Q\cos4\varphi$ with $Q=\avg{\cos4\theta}=0.090$ (dashed: the prediction). The
trapezoidal arc-length quadrature (blue) recovers a relatively flat amplitude, the
ripple shrinking from $0.090$ to $0.0035$. This is the projection used to build
the deficit.}
\label{fig:cos4phi}
\end{figure}

In figure~\ref{fig:cos4phi} we demonstrate both the error and its removal. 
By injecting a
unit $m{=}2$ at every orientation, the uniform mean ripples by
$Q = 0.090$, tracking $1+Q\cos4\varphi$ to the line, while the trapezoidal
quadrature recovers a relatively flat amplitude with a $\sim25\times$
smaller ripple.

In validation simulations of \S \ref{sec:validation}, we find a bias due to the
$\cos4\theta_w$ leak of the uniform quadrature's residual bias that changes
sign with WCS angle, $-0.0012 \to +0.0011$ ($\Delta_{\theta_w}\approx+0.0024$,
$\sim7\sigma$).  The trapezoidal quadrature flattens it to $+0.0004 \to
+0.0004$ ($\Delta_{\theta_w}\approx0$).

\section{The sky-angle deficit projection}
\label{sec:sky}

The second subtlety is which frame the harmonics are projected in. The
shear bias is a sky-frame contraction $\mathrm{Tr}[\mathbf{q}\,\mathbf{C}]$ between the object's
quadrupole and the noise covariance, so it is the sky-frame spin-2 that
must be cancelled, not the pixel-frame one. Under a non-conformal WCS (a pixel
grid that is sheared relative to the sky) the two differ.

For the projection we therefore bin and project in sky-frame angle. A pixel-frame
mode $e^{i\kvec\cdot\xvec}$ with sky coordinates $\mathbf u = M\xvec$ (where $M$
is the pixel$\to$sky Jacobian) has sky wavevector $M^{-T}\kvec$; normalizing out
the pixel scale (only the WCS shape matters),
\begin{equation}
  \kvec_{\rm sky} = \mathrm{scale}\cdot M^{-T}\,\kvec_{\rm pix},
\end{equation}
and the annulus binning and the $\cos m\theta/\sin m\theta$ projections use
$\kvec_{\rm sky}$. The Jacobian decomposes as $M = \mathrm{scale}\cdot \mathcal{R}(\theta)
\cdot S(g)$. A diagonal scale maps to the identity (a no-op) and a pure
rotation merely rotates $\theta$, in which case the $(c_m,s_m)$ is a no-op.
Only a genuine non-conformal shear changes the projection. Because the diagonal and
pure-rotation cases reduce to identity automatically, the WCS Jacobian
is all that is required. This sky-frame projection is the
analogue of doing the metacal shear in the world frame, and
of the sky-frame rotation of \S\ref{sec:rotfield}. Every spin-2 operation in the
code works in the sky frame.

\section{The stationary-noise limit}
\label{sec:synth}

The \fusion{} filter uses only second moments of the noise, and the deficit
$D_t$ is a non-negative power spectrum.  If the data noise is stationary
with a known spectrum $N(\kvec)$, there is no need to filter a real
realization. One can draw a single white field $\widehat w(\kvec)$ on the
shared grid and add $\mathcal F^{-1}[\sqrt{D^N_t}\,\widehat w]$ to each type,
the same $\widehat w$ for every type, where $D^N_t$ is the deficit of the
correlated transfer $\mathcal{P}_t(\kvec)\,N(S_t^{-1}\kvec)$, measured with
the noise kernel $\sqrt N$ in place of the delta (\S\ref{sec:filter}); for
white noise $D^N_t = D_t$.  This removes the need for the noise image, the sky
rotation, the second \mcalshort{} pass, $\mathcal{P}_t^{\rm rot}$, the division
and the cap. The variance increase is $1 + \avg{D^N_t}/\avg{\mathcal{P}_t
N}$, which is $1 + \avg{D_t}/\avg{\mathcal{P}_t}$ for white noise.  Where the
cap is inactive the two constructions add the same power in every mode, so for
Gaussian noise each type's output noise is statistically identical; the
correlation between types differs only through the phases of $K_t^{\rm rot}$,
which are below $0.02$\,rad in band for the scenes below.

We compared them on the scenes of \S\ref{sec:filter} with white Gaussian noise
($2000$ realizations, $48^2$ stamps).  For the round PSF they agree.  The
predicted $m{=}2$ anisotropy of the \code{1p} noise power falls from
$4.8\times10^{-2}$ to $\approx5\times10^{-4}$ in both, the variance factor is
$1.0428$ in both, and the $\code{1p}-\code{1m}$ noise is the same.  At
$g_1^{\rm psf}=0.15$ the cap binds on $30$--$50\%$ of the band and leaves
$3$--$7\%$ of the deficit unfilled, almost all of it $m{=}0$; both leave
$\approx0.5\%$ of the original $m{=}2$.  The Monte Carlo confirms both at the
statistical floor ($\approx10^{-2}$).  The synthesized field fills the common
level completely, so its variance factor is $1.712$ compared to the \fusion's
$1.669$, and in the stamp interior it adds correspondingly slightly more
$\code{1p}-\code{1m}$ variance ($2.4\times10^{-3}$ compared to
$\approx2\times10^{-3}$ per unit input variance).  Over the full stamp the
\fusion's $\code{1p}-\code{1m}$ variance is larger, because the sheared edge
of the finite correction stamp enters at full strength where $H_t\approx1$;
the synthesized field is stationary on the padded grid and has no edge.

This is a limiting case, not a replacement for the full \fusion{} correction.
Correlated noise is not necessarily the obstacle, since a known stationary $N$
enters through the kernel transfer. But $N$ must be known from the resampling
that produced it; it cannot be measured well enough from a single realization.
Furthermore, real coadd noise can be non-stationary (cell edges,
varying depth, masks), in which case there is no single $N$ to synthesize from;
the real realization represents that structure and $N$ (\S\ref{sec:filter}).
The synthesized form is useful for simulations and as a check on the \fusion{}
machinery.

\end{appendices}

\makeatletter
\let\jnl@style=\rmfamily
\def\ref@jnl#1{{\jnl@style#1}}
\newcommand\apj{\ref@jnl{ApJ}}
\newcommand\mnras{\ref@jnl{MNRAS}}
\newcommand\aap{\ref@jnl{AAP}}
\newcommand\jcap{\ref@jnl{JCAP}}
\newcommand\aj{\ref@jnl{AJ}}
\makeatother
\bibliographystyle{aasjournal}
\bibliography{references}

@ARTICLE{sheldon2017metacal,
  author = {{Sheldon}, Erin S. and {Huff}, Eric M.},
  title = {Practical Weak-lensing Shear Measurement with Metacalibration},
  journal = {The Astrophysical Journal},
  year = 2017,
  volume = {841},
  number = {1},
  pages = {24},
  doi = {10.3847/1538-4357/aa704b},
  eprint = {1702.02601},
  archivePrefix = {arXiv}
}

@ARTICLE{huff2017metacal,
  author = {{Huff}, Eric and {Mandelbaum}, Rachel},
  title = {Metacalibration: Direct Self-Calibration of Biases in Shear Measurement},
  journal = {arXiv e-prints},
  year = 2017,
  eprint = {1702.02600},
  archivePrefix = {arXiv}
}

@ARTICLE{sheldon2023rubin,
  author = {{Sheldon}, Erin S. and {Becker}, Matthew R. and {Jarvis}, Michael and {Armstrong}, Robert and {LSST Dark Energy Science Collaboration}},
  title = {Metadetection Weak Lensing for the Vera C. Rubin Observatory},
  journal = {The Open Journal of Astrophysics},
  year = 2023,
  volume = {6},
  pages = {17},
  doi = {10.21105/astro.2303.03947},
  eprint = {2303.03947},
  archivePrefix = {arXiv}
}

@ARTICLE{sheldon2026edges,
       author = {{Sheldon}, Erin},
        title = "{Edges In Coadded Images}",
      journal = {The Open Journal of Astrophysics},
         year = 2026,
        month = may,
       volume = {9},
        pages = {62176},
          doi = {10.33232/001c.162176},
archivePrefix = {arXiv},
       eprint = {2509.06929},
 primaryClass = {astro-ph.CO},
       adsurl = {https://ui.adsabs.harvard.edu/abs/2026OJAp....962176S}
}

@ARTICLE{galsim2015,
   author = {{Rowe}, B.~T.~P. and {Jarvis}, M. and {Mandelbaum}, R. and {Bernstein}, G.~M. and
    {Bosch}, J. and {Simet}, M. and {Meyers}, J.~E. and {Kacprzak}, T. and
    {Nakajima}, R. and {Zuntz}, J. and {Miyatake}, H. and {Dietrich}, J.~P. and
    {Armstrong}, R. and {Melchior}, P. and {Gill}, M.~S.~S.},
    title = "{GALSIM: The modular galaxy image simulation toolkit}",
  journal = {Astronomy and Computing},
archivePrefix = "arXiv",
   eprint = {1407.7676},
 primaryClass = "astro-ph.IM",
     year = 2015,
    month = apr,
   volume = 10,
    pages = {121-150},
      doi = {10.1016/j.ascom.2015.02.002},
   adsurl = {http://adsabs.harvard.edu/abs/2015A%26C....10..121R}
}

@misc{SRD,
  doi = {10.48550/ARXIV.1809.01669},
  url = {https://arxiv.org/abs/1809.01669},
  author = {Mandelbaum, Rachel and others},
  title = {The LSST Dark Energy Science Collaboration (DESC) Science Requirements Document},
  publisher = {arXiv},
  year = {2018},
  copyright = {arXiv.org perpetual, non-exclusive license}
}

@ARTICLE{mdet2020,
       author = {{Sheldon}, Erin S. and {Becker}, Matthew R. and {MacCrann}, Niall and {Jarvis}, Michael},
        title = "{Mitigating Shear-dependent Object Detection Biases with Metacalibration}",
      journal = {\apj},
         year = 2020,
        month = oct,
       volume = {902},
       number = {2},
          eid = {138},
        pages = {138},
          doi = {10.3847/1538-4357/abb595},
archivePrefix = {arXiv},
       eprint = {1911.02505},
 primaryClass = {astro-ph.CO},
       adsurl = {https://ui.adsabs.harvard.edu/abs/2020ApJ...902..138S}
}

@ARTICLE{GurvichMandelbaum2016,
       author = {{Gurvich}, Alex and {Mandelbaum}, Rachel},
        title = "{The impact of correlated noise on galaxy shape estimation for weak lensing}",
      journal = {\mnras},
         year = 2016,
        month = apr,
       volume = {457},
       number = {4},
        pages = {3522-3534},
          doi = {10.1093/mnras/stw174},
archivePrefix = {arXiv},
       eprint = {1509.00020},
 primaryClass = {astro-ph.IM},
       adsurl = {https://ui.adsabs.harvard.edu/abs/2016MNRAS.457.3522G}
}

@misc{sep,
  author       = {Kyle Barbary and
                  Kyle Boone and
                  Matt Craig and
                  Christoph Deil and
                  Benjamin Rose},
  title        = {kbarbary/sep: v1.0.2},
  month        = sep,
  year         = 2017,
  doi          = {10.5281/zenodo.896928},
  url          = {https://doi.org/10.5281/zenodo.896928}
}

@ARTICLE{Moffat1969,
   author = {{Moffat}, A.~F.~J.},
    title = "{A Theoretical Investigation of Focal Stellar Images in the Photographic Emulsion and Application to Photographic Photometry}",
  journal = {\aap},
     year = 1969,
    month = dec,
   volume = 3,
    pages = {455},
   adsurl = {http://adsabs.harvard.edu/abs/1969A%26A.....3..455M}
}

@ARTICLE{wldb2021,
       author = {{Sanchez}, Javier and {Mendoza}, Ismael and {Kirkby}, David P. and {Burchat}, Patricia R. and {LSST Dark Energy Science Collaboration}},
        title = "{Effects of overlapping sources on cosmic shear estimation: Statistical sensitivity and pixel-noise bias}",
      journal = {\jcap},
         year = 2021,
        month = jul,
       volume = {2021},
       number = {7},
          eid = {043},
        pages = {043},
          doi = {10.1088/1475-7516/2021/07/043},
archivePrefix = {arXiv},
       eprint = {2103.02078},
 primaryClass = {astro-ph.CO},
       adsurl = {https://ui.adsabs.harvard.edu/abs/2021JCAP...07..043S}
}

@ARTICLE{bj02,
       author = {{Bernstein}, G.~M. and {Jarvis}, M.},
        title = "{Shapes and Shears, Stars and Smears: Optimal Measurements for Weak Lensing}",
      journal = {\aj},
         year = 2002,
        month = feb,
       volume = {123},
       number = {2},
        pages = {583-618},
          doi = {10.1086/338085},
archivePrefix = {arXiv},
       eprint = {astro-ph/0107431},
 primaryClass = {astro-ph},
       adsurl = {https://ui.adsabs.harvard.edu/abs/2002AJ....123..583B}
}

\end{document}